\documentstyle[prb,aps,twocolumn,floats]{revtex}

\begin{document}
\input epsf
\draft
\wideabs{
\title {Critical currents, flux-creep activation energy and potential barriers for the 
        vortex motion from the flux creep experiments}

\author {I. L. Landau$^{1,2}$ H. R. Ott$^{1}$}
\address{$^{1}$Laboratorium f\"ur Festk\"orperphysik, ETH H\"onggerberg, 
CH-8093 Z\"urich, Switzerland}
\address{$^{2}$Kapitza Institute for Physical Problems, 117334 Moscow, 
Russia}

\date{\today}
\maketitle

\begin{abstract}
We present an experimental study of thermally activated flux creep in a superconducting 
ring-shaped epitaxial YBa$_{2}$Cu$_{3}$O$_{7-x}$ film as well as a new way of analyzing the 
experimental data. The measurements were made in a wide range of temperatures between 10 
and 83 K. The upper temperature limit was dictated by our experimental technique and at 
low temperatures we were limited by a crossover to quantum tunneling of vortices. It is 
shown that the experimental data can very well be described by assuming a simple thermally 
activated hopping of vortices or vortex bundles over potential barriers, whereby the hopping 
flux objects remain the same for all currents and temperatures. The new procedure of data 
analysis also allows to establish the current and temperature dependencies of the flux-creep 
activation energy $U$, as well as the temperature dependence of the critical current $I_{c}$, 
from the flux-creep rates measured at different temperatures. The variation of the activation 
energy with current, $U(I/I_{c})$, is then used to reconstruct the profile of the potential 
barriers in real space. 
    
\end{abstract}
}    

\section {Introduction}

Investigations of the flux-creep process in type-II superconductors reveal important 
information about the interaction of vortices with pinning centers and among the vortices 
themselves. Studies of this type are especially rewarding for high temperature superconductors 
(HTSC's) because in these materials a particularly rich variety of features of the vortex 
state has been established. In principle, the analysis of flux-creep data obtained at 
different temperatures permits to establish the dependence of the flux-creep activation 
energy $U$ on the current density $j$ and on temperature $T$. Different scaling procedures 
have been developed and used in order to deduce this information.$^{1-19}$ However, the 
interpretation of the experimental results is rather complicated because the suggested 
procedures involve many parameters which are not a priori known. Actually, there is no way 
to deduce all the parameters from the experimental data alone and some additional assumptions 
have to be made. The lacking input is usually provided by invoking different theoretical 
models and therefore the final result naturally depends on the particular chosen model. In 
many cases, different models have been employed to interpret data obtained from the same 
kind of samples, resulting, for example, in rather differing $U(j)$ curves. This is why, in 
spite of the extensive literature on this subject, the available information following 
from the analyses of the experimental data is still, to a certain degree, inconclusive and 
often controversial.

Recently we have proposed a new approach for analyzing the flux-creep rates in HTSC's.$^{20}$ 
This approach is based on a few basic assumptions and it essentially consists in merging the 
experimental voltage-current ($V$-$I$) characteristics of one sample, obtained at different 
temperatures, using their shape as the key to deduce the scaling parameters. It has been 
demonstrated that this approach works rather well for $V$-$I$ characteristics of a ring-shaped 
film of YBa$_{2}$Cu$_{3}$O$_{7-x}$ (YBCO) in the temperature range between 10 and 60 K, the 
temperature interval covered in the work previously published in Ref. 20. The proposed scaling 
procedure permits to establish the dependence of the flux-creep activation energy $U$ on the 
normalized current density $j/j_{c}$, where $j_{c}$ is the critical current density, as well 
as the temperature dependence of the critical current, directly from the $V$-$I$ characteristics 
of the sample in the flux-creep regime. The main goal of the present work was to test whether 
the same procedure may also be applied successfully at temperatures closer to $T_{c}$. 
Therefore we have rearranged the experimental setup such as to allow an extension of the 
measurements up to 83 K. This extension of the measurements to higher temperatures is 
important because it provides information about $U(j/j_{c})$ for low values of $j/j_{c}$. In 
this way the covered range of currents has been extended down to $j/j_{c}\approx 0.05$, 
approximately an order of magnitude lower than $j/j_{c}\approx 0.4$, reached in Ref. 20.

In addition to monitoring the current decay in zero external field as described in Ref. 20, 
we have extended the data base by measuring the flux-creep rates also in external magnetic 
fields of 0.3 and 1 kOe.

\section {Experimental}

\subsection{Sample and measurements}

The experiments have been made using a ring-shaped epitaxial YBCO film with a superconducting 
critical temperature $T_{c} = 87.5$ K. The external diameter of the ring is 10 mm and its 
width is approximately 2 mm. The film thickness is about 0.3 $\mu$m. The resistive transition 
to  superconductivity of the sample is shown in the inset of Fig. 3. More details about the 
sample and the basic experimental set-up can be found elsewhere.$^{20-22}$ 

For the present study we intended, as mentioned above, to extend the measurements to 
temperatures as close to $T_{c}$ as possible. The main technical obstacle is the very strong 
temperature dependence of the flux-creep rate at temperatures close to $T_{c}$, asking for a 
high stability and accuracy of the temperature control for obtaining reliable data. The 
desired temperature stability has been achieved by using a Platinum resistance thermometer 
for temperatures exceeding 30 K. With this temperature sensor the computer-based temperature 
controller provided a temperature stability of $\pm 1$ mK. For lower temperatures we used a 
diode thermometer providing a temperature stability of $\pm 30$ mK, sufficient in this 
temperature range.

Step-like changes of the external magnetic field $H$, oriented perpendicularly to the ring 
plane, were used to induce an electrical current in the ring. Three different procedures were 
employed, i.e., (i) switching off a field $H = 1$ kOe to $H_{e} = 0$, (ii) switching the 
field to a value of $H_{e} = 1$ kOe, and (iii) switching the field to a value of $H_{e} = 0.3$ 
kOe. In the last two cases both positive ($0 \rightarrow H_{e}$) and negative 
($H_{0} \rightarrow H_{e}$) field steps with $H_{0} = 1.7$ kOe for the case $H_{e} = 1$ kOe 
and $H_{0} = 0.6$ kOe for $H_{e} = 0.3$ kOe were made.

\begin{figure}[h]
 \begin{center}
  \epsfxsize=0.9\columnwidth \epsfbox {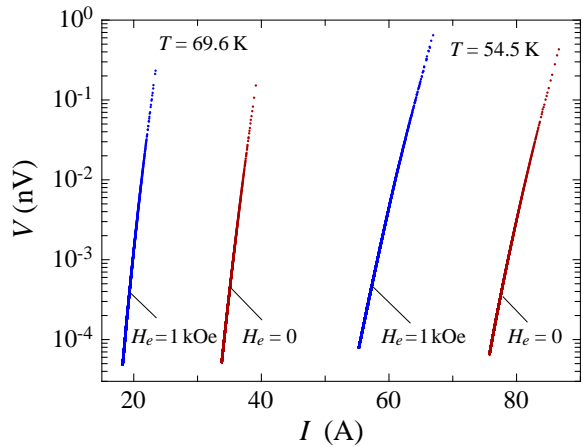}
  \caption{Examples of $V$-$I$ characteristics at two different temperatures.}
 \end{center}
\end{figure}
After these stepwise variations of the external field, the magnetic induction in the ring 
cavity $B_{i}$ was monitored as a function of time $t$. For this purpose a LakeShore 450 
Gaussmeter with a standard cryogenic Hall probe was used. From the $B_{i}(t)$ data, the 
current decay curves $I(t)$ may be calculated straightforwardly, taking into account the 
position of the Hall probe inside the ring cavity. Using the $I(t)$ data, the voltage around 
the ring sample can be calculated via $V=LdI/dt$, where $L \approx 8$ nH is the sample 
inductance. The primary experimental data can thus easily be converted into $V$-$I$ 
characteristics of the sample. Examples of collected $V$-$I$ curves are shown in Fig. 1.

In this kind of experiments it is very important to make sure that the current density in the 
sample, induced by the magnetic field step, is high enough to create the critical state 
throughout the sample. In this case the experimental results, represented as $V$-$I$ curves, 
are practically independent of the magnitude of the field step as well as the magnetic 
history of the sample. We note that for $H_{e} = 0$ and $H_{e} = 1$ kOe, this was indeed the 
case for the whole covered temperature range. For $H_{e} = 0.3$ kOe, however, the step 
magnitude was insufficient at low temperatures and the measurements were feasible at 
$T \geq 70$ K only.

\subsection {Heating effects}

In this kind of experiments it is also essential to avoid an overheating of the sample via 
Joule heating caused by the induced current. In the flux creep regime the dissipation power 
is negligibly small and there is no overheating. However, during the abrupt change of the 
external magnetic field the induced transient current may be higher than the critical current 
$I_{c}$¥ and therefore, the heating effects may be considerable. During the time period of the 
magnetic field step, the voltage around the sample may be estimated as
\begin{equation}
V=-{1 \over c}{{d\Phi } \over {dt}},
\end{equation}
where $c$ is the speed of light and $\Phi$ is the magnetic flux inside the ring cavity. In our 
experiments the duration of the field variation was of the order of 50 ms. This implies a 
voltage $V \sim 100$ $\mu$V, which is more than 5 orders of magnitude higher than typical 
voltages in the flux creep regime.

Our thin film sample has a low heat capacity and it is in good thermal contact with the 
substrate. Therefore the thermal equilibrium should be restored much quicker than the time 
given by the delay of a few seconds between the field step and the beginning of monitoring 
$I(t)$. In this case any overheating effects are negligible in a large part of the covered 
temperature range. However, at temperatures close to $T_{c}$, the situation is quite 
different. In this case, during the field step, the sample may be heated to above $T_{c}$ 
and the current, induced by the field step, may decay considerably before superconductivity 
in the sample is restored. If the current decays too strongly, the resulting current density 
may not be sufficient for the creation of the critical state in the sample and the flux-creep 
data will be distorted. 

In order to illustrate this problem we show the current decay curves for 3 temperatures in 
the high temperature range in Fig. 2. The curve corresponding to $T = 77.7$ K demonstrates 
a slight upward curvature which is typical for flux-creep behavior. At $T = 78.7$ K the 
\begin{figure}[h]
 \begin{center}
  \epsfxsize=0.9\columnwidth \epsfbox {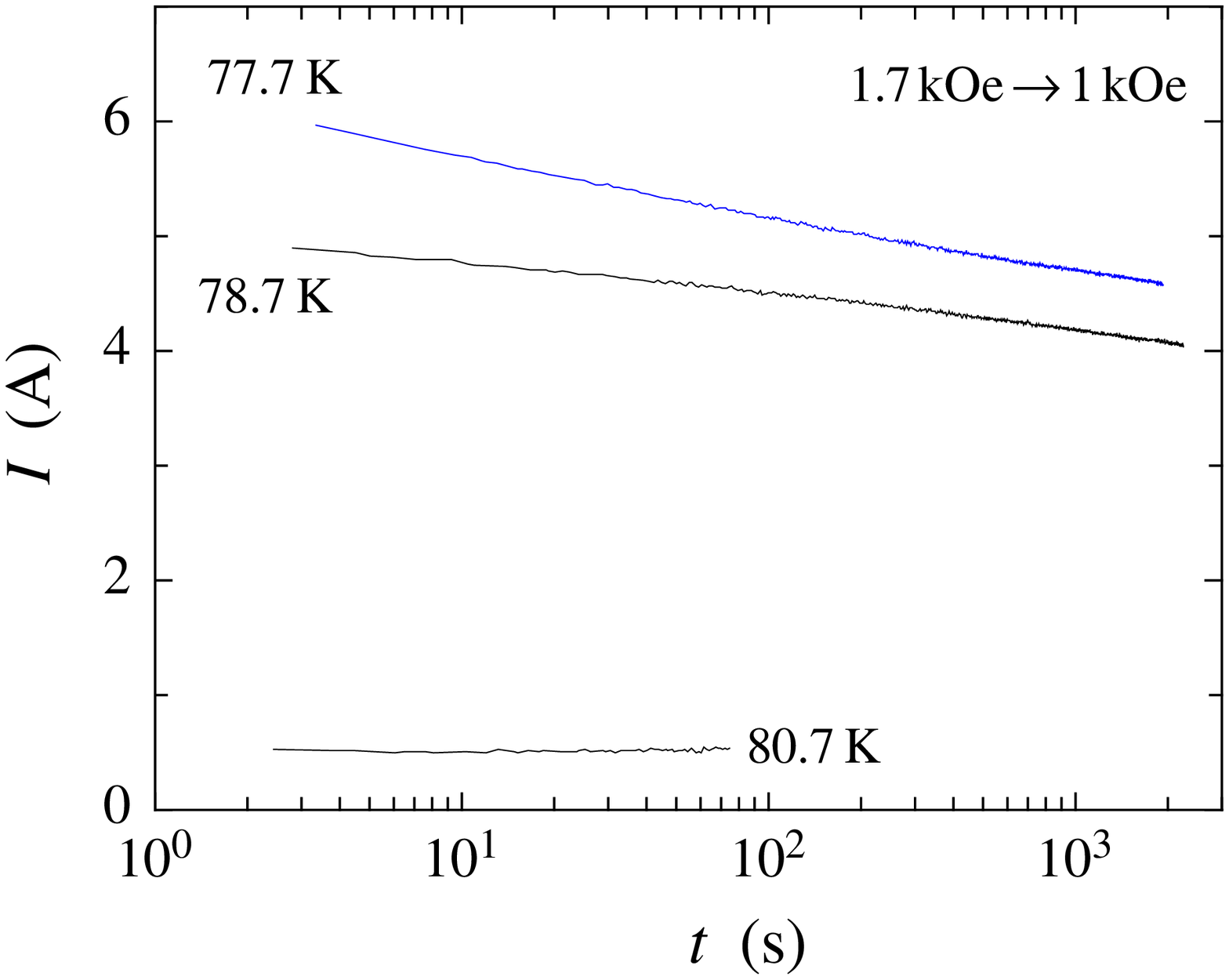}
  \caption{Variations of the electric current $I$ in the sample as a function of time after 
           magnetic field step at three temperatures in the high-temperature range.}
 \end{center}
\end{figure}
situation is already different. The curvature is of opposite sign, indicating that the 
current density was insufficient to create the critical state. At $T = 80.7$ K the current 
is close to zero from the very beginning and the flux creep phenomenon is no longer reflected 
in the $I(t)$ curve. In this way, overheating limits the temperature range where useful 
experiments of this kind may be made. Different field steps have different upper temperature 
limits. In our work the lowest limit is attained when the external magnetic field is switched 
to 1 kOe. As is illustrated in Fig. 2, in this case meaningful measurements are not possible 
above $T \approx 78$ K. For $H_{e} = 0$, the limiting temperature was about 81 K, whereas for 
$H_{e} = 0.3$ kOe, measurements up to $T \approx 83$ K were possible.

\subsection {Magnetic induction in the sample}

The important parameter in the flux-creep process is the magnetic induction $B$ in the bulk 
of the sample. The magnetic induction fixes, for instance, the vortex density. In our 
experiments we did not measure $B$ and there is no way to estimate it accurately. After the 
magnetic field step has been applied, $B$ must adopt a value somewhere between those that 
correspond to the initial and the final values of $H$. This is, of course, a very rough 
estimate, especially in the case when the external field is switched off ($H_{e} = 0$). A 
redistribution of the magnetic induction in the sample only occurs during the field step and 
then, in the flux-creep regime, $B$ remains practically constant in time.

For $H_{e} = 0$ the magnetic induction is due to the remanent magnetization. At low 
temperatures, where the critical current density is practically temperature 
independent,$^{21}$ $B$ should be independent of temperature as well. At higher temperatures 
however, $B$ decreases with increasing temperature, tending to zero at $T = T_{c}$. This uncertainty in 
$B$ greatly complicates the interpretation of the experimental data at temperatures close to 
$T_{c}$.

If $H_{e} \neq 0$, the magnetic induction $B$ is larger than the value corresponding to 
$H_{e}$ for the negative field step and by about the same amount smaller for the positive 
step. In this case it is more appropriate to use the data averaged for two different field 
steps rather than the individual results. The averaged results thus correspond to a 
temperature independent $B = 0.3$ kG or $B = 1$ kG for $H_{e} = 0.3$ kOe and $H_{e} = 1$ kOe, 
respectively. The averaging also considerably reduces the experimental errors, as it is
discussed in more details in Ref. 21. This is why for $H_{e} \neq 0$ we present only the 
averaged results.

\section {Experimental results and analysis}
\begin{figure}[h]
 \begin{center}
  \epsfxsize=0.9\columnwidth \epsfbox {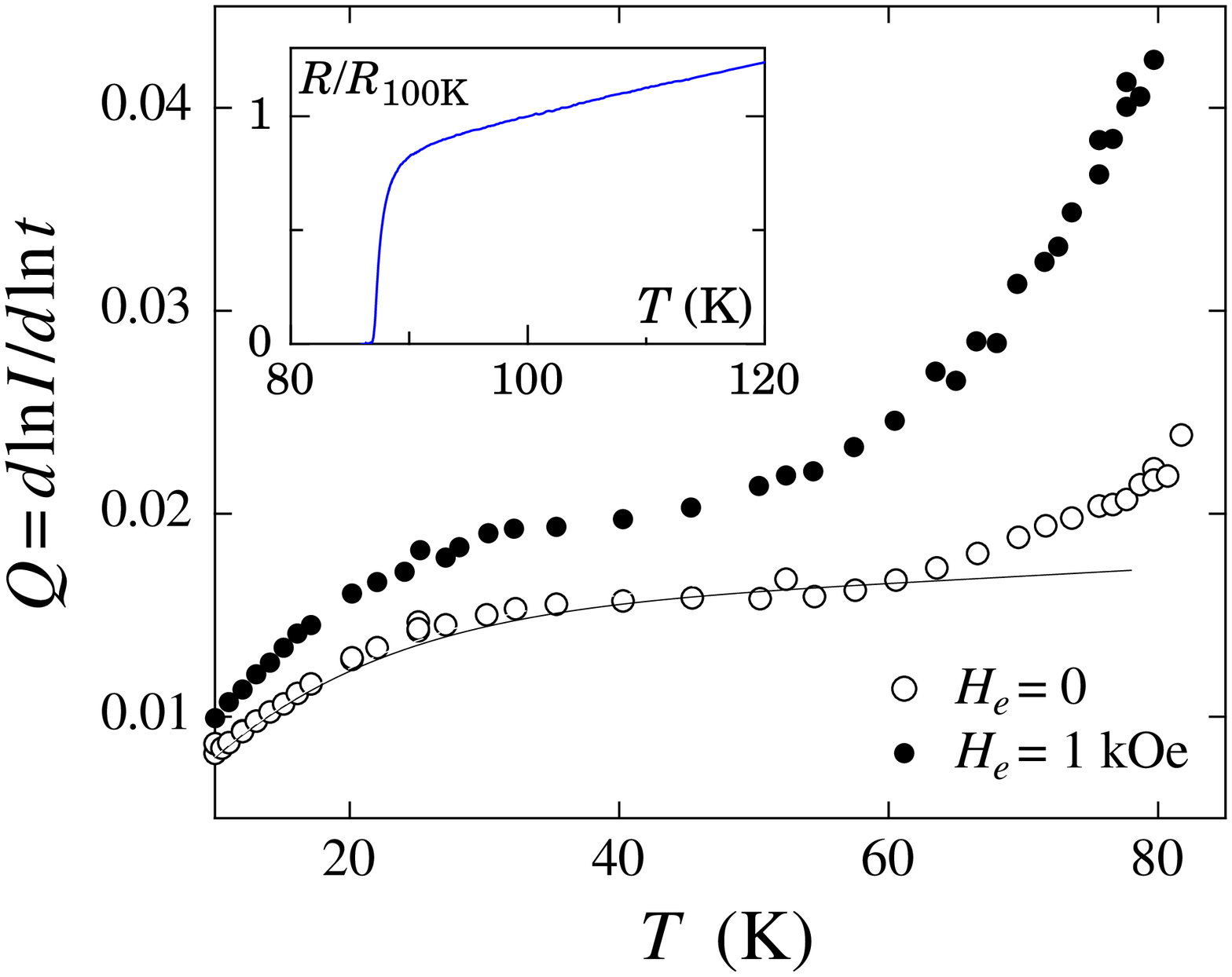}
  \caption{The normalized relaxation rate $Q=d\ln I/d\ln t$, obtained for $H_{e} = 0$ and 
           1 kOe, as a function of temperature. The symbols represent the experimental data. 
		   The solid line represents $Q(T)$ calculated for the potential profile $u(x)$, 
		   shown in Fig. 11, for $V = 0.1$ nV. The temperature dependence of the critical 
		   current was neglected in the calculations. The calculation procedure is described 
		   in Section IV. The inset shows the resistive superconducting transition of the 
		   sample.}
 \end{center}
\end{figure}
One of the distinct features of the magnetization relaxation in YBCO compounds is the 
existence of a plateau in the temperature dependence of the normalized relaxation rate 
$Q = d \ln M_{irr} / d \ln t$, where $M_{irr}$ is the non-equilibrium magnetization of the 
sample. Such plateaus with approximately the same values of $Q$ have been observed for 
different kinds of YBCO samples, including epitaxial films.$^{12,23,24}$ Fig. 3 displays 
the temperature dependencies of $d \ln I / d \ln t$, which is an exact equivalent of $Q$, 
for our sample. We note the typical plateau in the intermediate temperature range, obviously 
more pronounced in the case of $H_{e} = 0$. It should be noted that the $\ln I$ versus 
$\ln t$ curves are not exactly straight lines. In this case, the value of $Q$ depends on 
the time $t$, at which the derivative is taken. We have chosen the value of $t = 400$ s to 
evaluate $Q$ for the data presented in Fig. 3. 

The normalized relaxation rate may also be defined as $Q = - d \ln I / d \ln V$, which is 
equivalent to $Q = d \ln I / d \ln t$ if both derivatives are established at same value of 
current. In our case the chosen time corresponds to the voltage $V = 0.01$ nV, which is 
practically independent of temperature for $T \leq 40$ K. At higher temperatures, however, 
this voltage decreases with increasing temperature.

\subsection {Scaling procedure}

If the current $I$ in the sample is less than its critical value $I_{c}$, all vortices are
pinned and their motion occurs only due to either thermally activated hopping over the 
potential barriers or via quantum tunneling. The latter mechanism is dominant at low 
temperatures. For our sample the crossover from thermal activation to quantum tunneling 
occurs at $T \approx 10$ K. The low temperature features have thoroughly been investigated 
in Ref. 21 and in the present work we consider thermal activation only. Assuming that the 
change of the magnetic flux in the ring cavity is due to thermally activated hopping of 
vortices in the sample, i.e., due to flux creep, the voltage around the sample is
\begin{equation}
V=V_0\exp \left( {-{U \over {k_BT}}} \right),
\end{equation}
where
\begin{equation}
V_0={{\nu _0l_{hop}BL_{creep}} \over c}.
\end{equation}
Here $U$ is the flux-creep activation energy, $k_B$ is the Boltzmann constant, $\nu_0$ is an 
attempt frequency of the vortices to cross the potential barrier, $l_{hop}$ is the vortex 
hopping distance, and $L_{creep}$ is that length of the sample, which contributes to the flux 
creep.$^{25}$ $L_{creep}$ is difficult to evaluate, however, it does not depend on current, 
temperature or external magnetic field. An electrical current creates a Lorentz force acting 
on the vortices $F_{L}$ which tilts the potential profile, thus reducing the potential 
barriers for the vortex motion. 

Using Eq. (2), the flux-creep activation energy may be expressed as
\begin{equation}
U(I)=-k_{B}T[\ln V(I)-\ln V_0].
\end{equation}
The value of the current at which $U(I)$ vanishes is a formal definition of the critical 
current $I_{c}$. According to Eq. (4), the parameter $V_0$ is equal to $V$ at $I = I_{c}$. 

Eq. (4) offers a way to extract $U(I)$ from experiment. Unfortunately, the experimental data 
sets of $V(I)$ at different temperatures cover only a very narrow range of currents. An 
additional complication in using of Eq. (4) for evaluating $U(I)$ is that neither $V_{0}$ nor 
$I_{c}$ are a priori known. In order to expand the available current range, numerous attempts 
to scale the data sets obtained at different temperatures have been made.$^{1-19}$ The most 
reliable procedure is provided by the Maley method,$^{1}$ which does not invoke any a priori 
assumptions. This method, however, is only applicable if both the flux-creep activation 
energy and the critical current are temperature independent. In this case, Eq. (4) implies 
that the $V$-$I$ curves for different temperatures, plotted as $T \ln V$ versus $I$, represent 
different parts of the same $U(I)$ curve, but are shifted vertically with respect to each 
other. The application of Maley's method to experimental $V$-$I$ curves provides a direct way 
to evaluate $\ln V_{0}$ and to determine $U(I)$. In general, however, the activation energy 
$U$ and the critical current $I_{c}$ are temperature dependent and the scaling of the 
flux-creep data turns out to be a rather complicated problem. 

With this in mind we have recently developed a new approach of scaling the $V$-$I$ curves in 
the flux creep regime.$^{20}$ This new procedure is based on merging the experimental $V$-$I$ 
curves, using their curvature for establishing the scaling parameters. The main assumption 
is that the flux creep is due to thermally activated hopping of vortices or vortex bundles 
over potential barriers and that these hopping flux objects remain the same for all 
temperatures and currents. This assumption implies that an electric current does not alter 
the interaction of vortices with the pinning centers and therefore the potential profile for 
a non-zero current is obtained by a linear superposition of the zero-current potential 
profile and a term arising from the Lorentz force. Below we briefly discuss the essential 
consequences of this assumption, more details may be found in Ref. 20.

We start by considering the profile $u(x)$ of the potential energy for the vortex motion in 
the vicinity of one of the potential wells. The $x$-axis coincides with the direction of the 
flux motion and $x = 0$ is chosen at the inflection point of the $u(x)$-function. The Lorentz 
force acting on vortices can be written as
\begin{equation}
F_L=j{{n\delta \Phi _0} \over c},
\end{equation}
where $j$ is the current density, $n$ is the number of vortices in the moving vortex bundle, 
$\delta$ is the sample thickness and $\Phi_0$ is the magnetic flux quantum. Taking into account 
the Lorentz force, we get the potential profile for a non-zero current as
\begin{equation}
u(x,j)=u(x,0)-xF_L.
\end{equation}
The important implication of Eq. (6) is that for any smooth function $u(x)$ the distance 
between the bottom of the well and the adjacent potential maximum along the positive $x$-axis 
decreases with increasing current and vanishes at $I = I_{c}$. This situation, first pointed 
out by Beasley et al.,$^{28}$ implies that the flux-creep activation energy is a non-linear 
function of current for any reasonable shape of the potential profile.$^{26}$ This 
non-linearity of $U(I)$ results in an upward curvature of the current decay curves, which may 
be seen in Fig. 2, and in a downward curvature of the $\log V$-$I$ curves depicted in Fig. 1.

The critical current density is reached if the potential barriers vanish. According to Eqs. (5) 
and (6), this results in
\begin{equation}
j_c={{cu'_c} \over {n\delta \Phi_0}}
\end{equation}
where $u'_c$ is the maximum value of $du(j=0)/dx$. This value is reached at the inflection 
point, i.e., at $x = 0$. In the following we assume that not the shape, but only the amplitude 
of the $u(x)$-function is temperature dependent, i.e., 
\begin{equation}
u(x)=U_0(T)f(x).
\end{equation}
where $U_0(T)$ is the temperature dependent amplitude of the $u(x)$-function. Eq. (8) 
represents the second assumption, on which our scaling procedure is based. Using both our 
assumption that the structure of the hopping flux object is independent of temperature and 
Eq. (8), the flux-creep activation energy may be written as
\begin{equation}
U(I,T)=U_0(T)Y(I/I_c),
\end{equation}
where the function $Y$ depends only on the ratio $I/I_c$.$^{20}$ By comparing Eq. (9) with 
Eqs. (7) and (8), it is obvious that both the critical current and the activation energy 
exhibit the same temperature dependence, given by the function $U_0(T)$.

At currents close to $I_{c}$ only a small part of the $u(x)$-function in the vicinity of the 
inflection point is essential in the formation of potential barriers. Since $u(x)$ is 
virtually a linear function in this region, the validity Eq. (8) is practically obvious. At 
lower currents, however, the flux-creep activation energy is determined by the features of 
$u(x)$ far away from the inflection point and the applicability of Eqs. (8) and (9) is 
difficult to justify a priori. As will be shown below, the analysis of our experimental 
results strongly indicates that the conditions expressed in Eqs. (8) and (9) are actually 
valid for a very wide range of currents down to $I\sim 0.1I_c$.

In Ref. 20 it was shown that, if the flux-creep activation energy may indeed be written as a 
product of a temperature and a current dependent term, the following transformation 
\begin{equation}
T_0\ln V(I/i,T_0)={{T\ln V(I,T)} \over i}+AT_0,
\end{equation}
where
\begin{equation}
A=\left( {1-{T \over {iT_0}}} \right)\ln V_0,
\end{equation}
may be used to merge the $V$-$I$ curves at different temperatures into a single master curve. 
Here, $i = I_{c}(T)/I_{c}(T_{0}) = U_{0}(T)/U_{0}(T_{0})$ and $A$ are the scaling parameters, 
and $T_{0}$ is some arbitrary chosen temperature within the investigated temperature range. 
The resulting master curve represents the current dependence of $T \ln V$ at $T = T_{0}$, as 
if $V(I)$ could actually be measured over this extended range of currents at this single 
temperature. For each temperature the values of $i$ and $A$ can be found from the condition 
that the overlapping parts of the $T \ln V$ versus $I$ curves for the adjacent temperatures 
match each other. It is important to recall that in this procedure we do not use the relation 
between $i$ and $A$ given by Eq. (11), but consider them as independent fitting parameters. 
Eq. (11) is used retrospectively to check the validity of our approach.

\begin{figure}[h]
 \begin{center}
  \epsfxsize=0.9\columnwidth \epsfbox {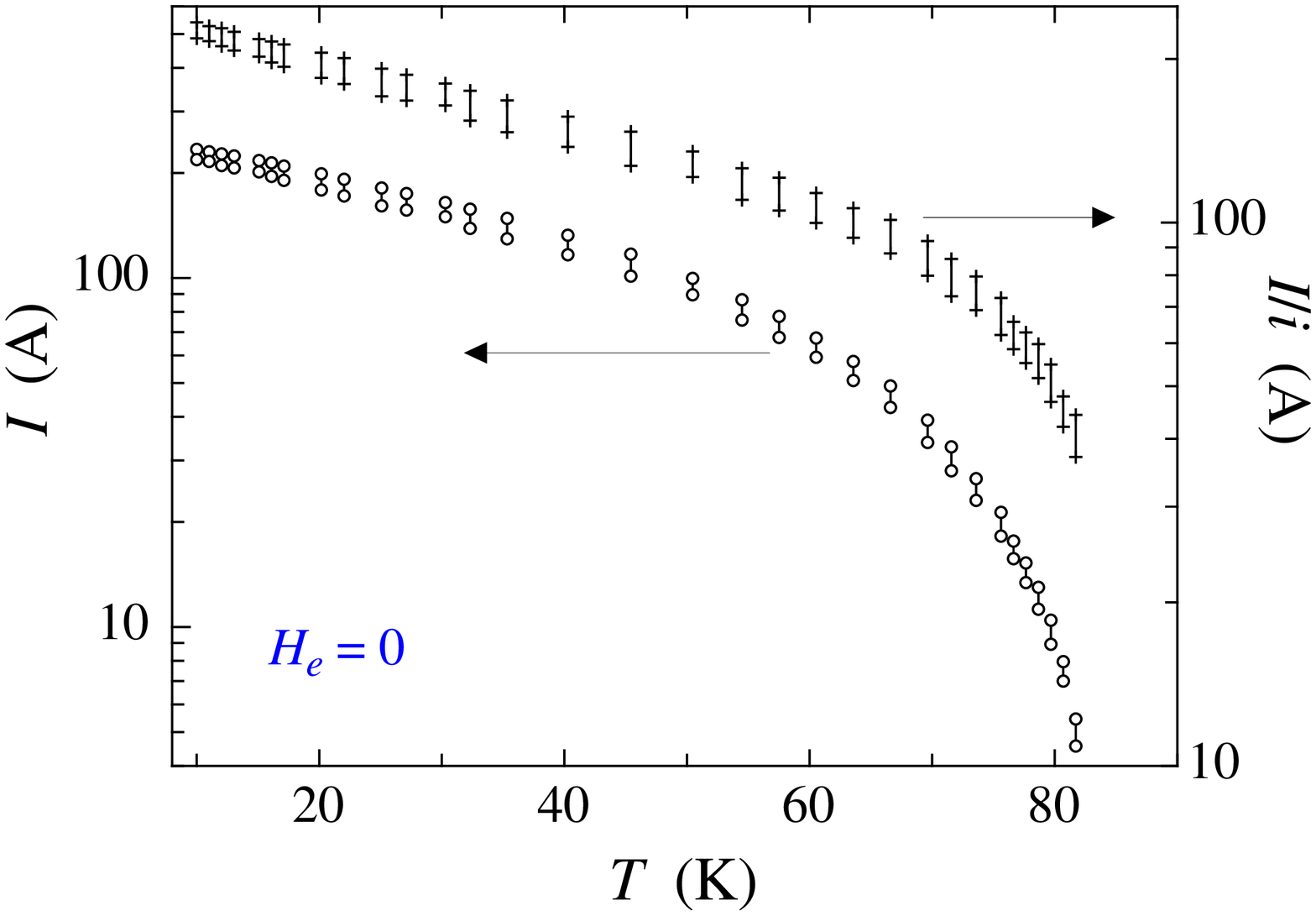}
  \caption{Covered current ranges at many different temperatures and at $H_{e}¥ = 0$. The 
           right hand scale is a renormalized current scale.}
 \end{center}
 \end{figure}
A successful application of the proposed scaling procedure demands that the current decay 
measurements are made at temperatures separated by sufficiently small intervals, such as to 
ascertain a considerable overlap of the $V$-$I$ curves for neighboring temperatures. Fig. 4 
displays the full set of the current ranges covered by the $V$-$I$ curves at each temperature 
and $H_{e} = 0$. The left vertical scale denotes the absolute values of the current, while 
the right one represents the normalized values. The latter set of data demonstrates that in 
all cases the overlap of the $V$-$I$ curves were sufficient to ensure a satisfying accuracy of 
the scaling procedure.

\subsection {Results for $H_{e} = 0$ and $H_{e} = 1$ kOe}

In this section we present the results of the scaling procedure for $H_{e} = 0$ and 
$H_{e} = 1$ kOe. The measurements for these two cases were made down to $T = 10$ K. The case 
of $H_{e} = 0.3$ kOe, which could only be studied at $T\geq 70$  K, will be discussed in the 
next section.

We have applied the scaling procedure according to Eq. (10) to our experimental $V$-$I$ curves 
and the corresponding master curves are shown in Fig. 5. As may be seen, the outlined scaling 
procedure provides the corresponding master curves by a practically perfect alignment of the 
$T \ln V$ versus $I$ curves obtained at different temperatures, as it is emphasized in the 
inset of Fig. 5. 

\begin{figure}[t]
 \begin{center}
  \leavevmode
  \epsfxsize=0.9\columnwidth \epsfbox {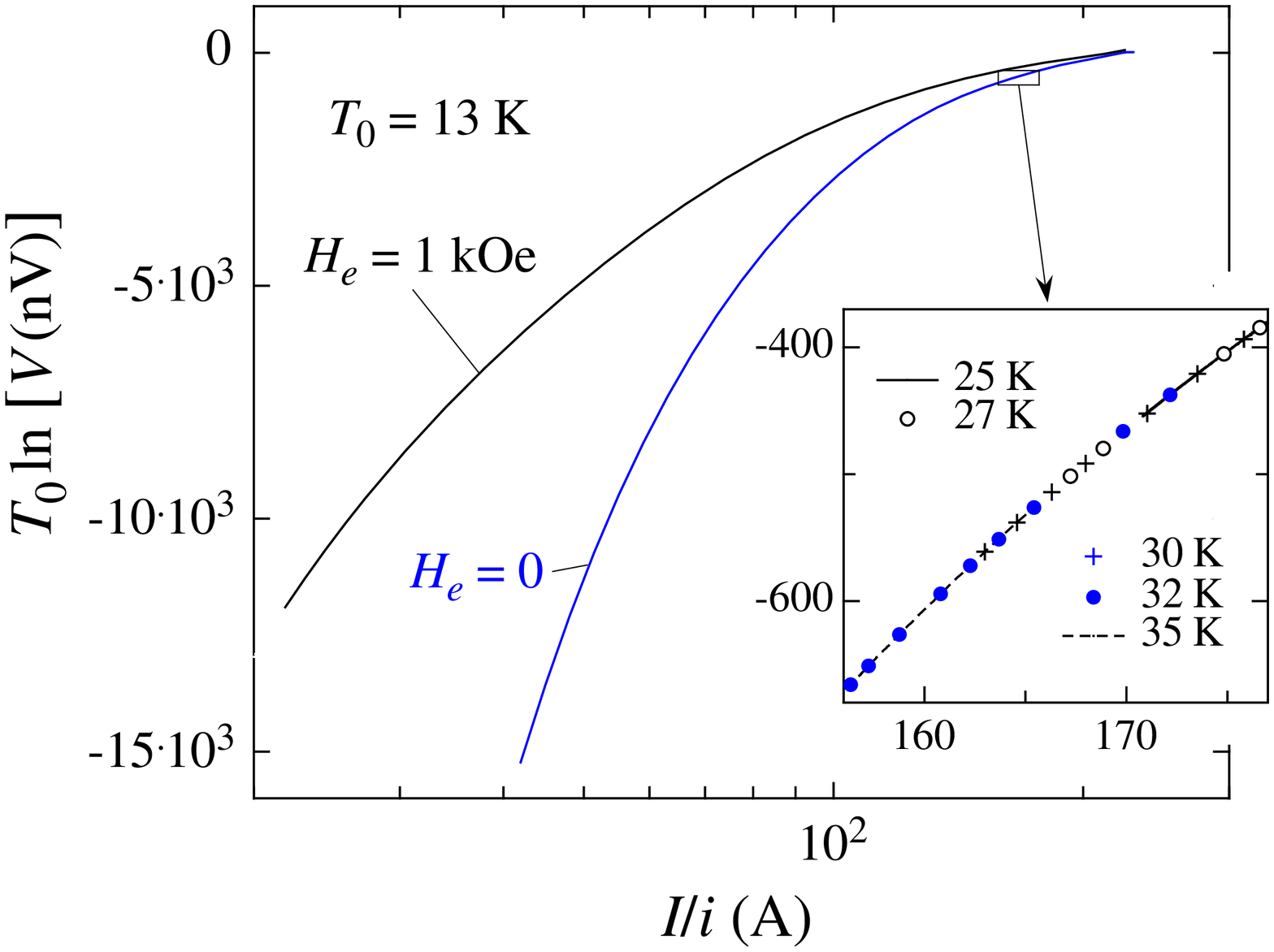}
  \caption{Results of the scaling procedure in the form of $T_{0}\ln V(T_{0})$ versus 
           $I/i$ with $T_{0}¥ = 13$ K. The inset shows, on linear scales, the small part of 
		   the curve for $H_{e}¥ = 0$ which is indicated by the rectangle in the main figure. 
		   For clarity only very few points for each temperature are displayed.}
  \label{TOlnV(T0)}
 \end{center}
\end{figure}
\begin{figure}[h]
 \begin{center}
  \leavevmode
  \epsfxsize=0.9\columnwidth \epsfbox {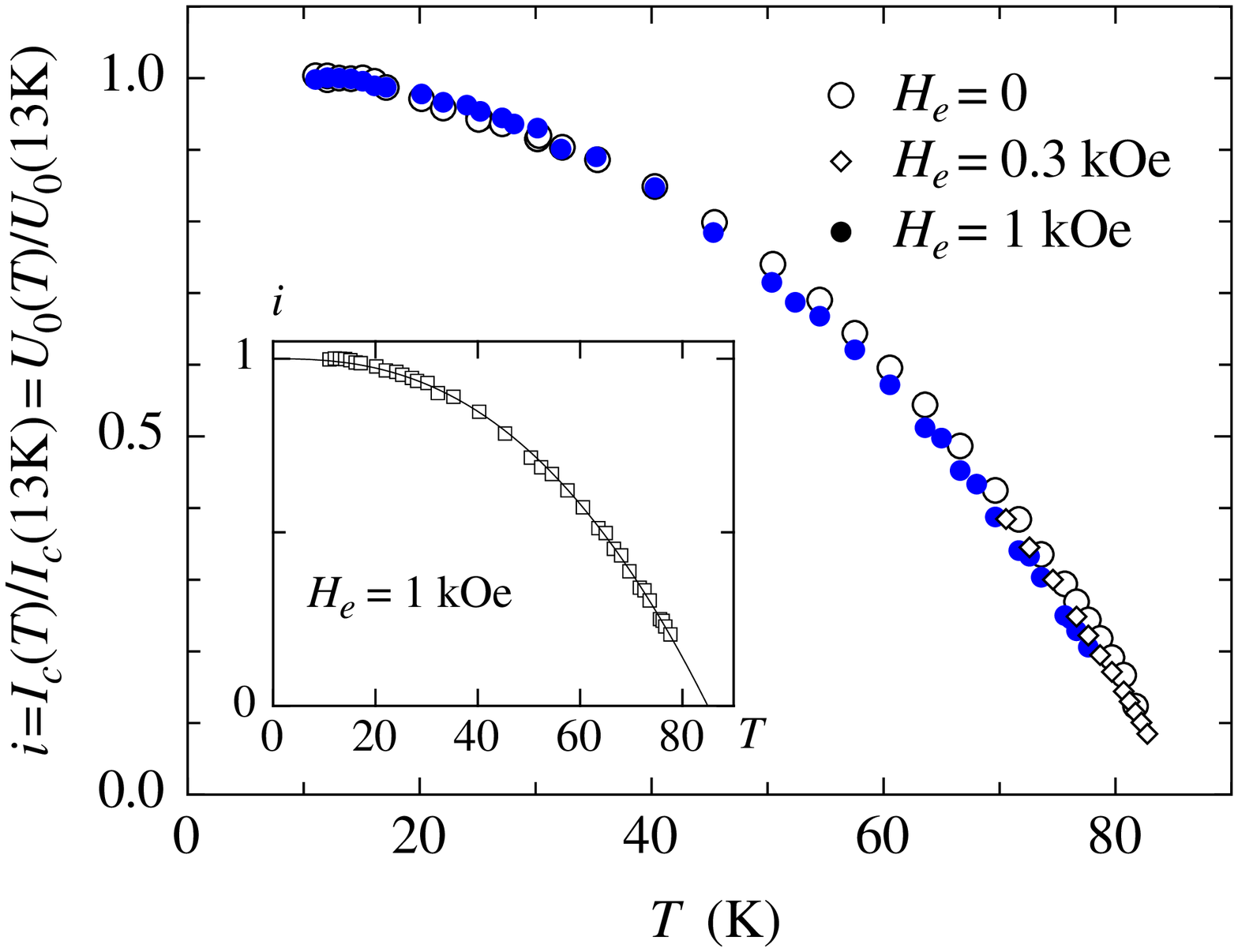}
  \caption{The scaling parameter $i$ as a function of temperature. The inset shows, as the 
           solid line, a fit to the data (open squares) using Eq. (12).}
  \label{ii(T)}
 \end{center}
 \end{figure}
Fig. 6 shows the temperature dependence of the scaling parameter $i$. In our approach, this 
plot represents the temperature dependence of the normalized critical current. Although in 
Fig. 5 the $T_{0} \ln V$ versus $I/i$ curves for $H_{e} = 0$ and $H_{e} = 1$ kOe are rather 
different, the respective $i(T)$ curves for these two cases almost coincide. The small 
difference between the two sets of data at higher temperatures is to be expected, when taking 
into account the suppression of the critical current by the external magnetic field. It is the 
magnetic induction $B$ in the sample which dictates the value of the critical current. As has 
already been mentioned, for $H_{e} = 0$ we are dealing with the temperature dependent remanent 
magnetization, and therefore, $B$ is not constant across the covered temperature range, but 
tends to zero at $T = T_{c}$. Somewhat simpler is the case where $H_{e} = 1$ kOe, corresponding 
to a temperature independent $B = 1$ kG. For this situation we note that $i(T)$ may very well 
be approximated by a simple power law 
\begin{equation}
i(T)=1-(T/T_{dp})^\mu ,
\end{equation}
across the whole covered temperature range. This is illustrated in the inset of Fig. 6, where 
the solid line represents the fit using the function of Eq. (12), with the fit parameters 
$\mu = 2.5 \pm 0.01^{27}$ and $T_{dp} = 84.95 \pm 0.05$ K$^{27}$. Quite surprisingly, the 
value of the exponent turns out to be exactly 5/2. Eq. (12) implies a linear dependence of 
the critical current on temperature near $T_{dp}$. 

 \begin{figure}[h]
 \begin{center}
  \epsfxsize=0.9\columnwidth \epsfbox {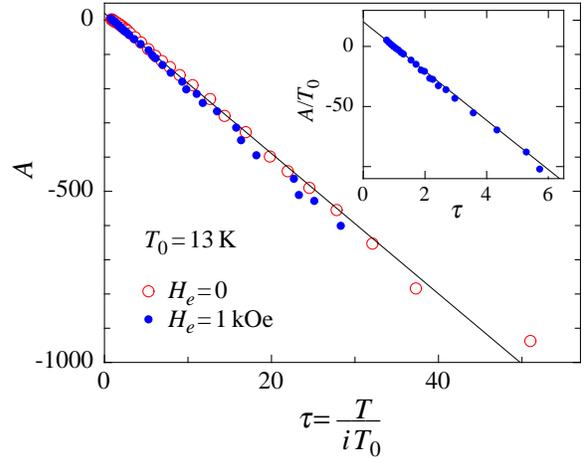}
  \caption{The parameter $A$ as a function of the normalized temperature $\tau = T/iT_{0}$ 
           with $T_{0}¥ = 13$ K. The straight line is drawn according to Eq. (11) with 
		   $\ln [V_0($nV$)]=20.5$. The inset shows the low temperature part of the plot on 
		   expanded scales.}
 \end{center}
\end{figure}
Next we consider the temperature dependence of the scaling parameter $A$. According to Eq. 
(11), $A$ depends on the ratio $T/i$ rather than the temperature alone. In Fig. 7, $A$ is 
plotted as a function of $\tau = T/iT_{0}$. If the temperature dependence of $\ln V_{0}$ is 
negligible and our procedure is self-consistent, we expect the data to lie on a straight line. 
Although, according to Eq. (3), $V_{0}$ is proportional to the temperature dependent attempt 
frequency, it enters Eq. (11) only as $\ln V_{0}$ and therefore, the resulting curve is 
expected to deviate rather weakly from linearity. This is indeed the case, as may be seen in 
Fig. 7. It is also remarkable that the data for $H_{e} = 0$ and $H_{e} = 1$ kOe are rather 
close to each other across the entire covered temperature range. This is to be expected, 
however. In our model the only difference between these two cases is the different values of 
the magnetic induction $B$ in the sample, which enters Eq. (11) as $\ln V_{0}$ (see Eq. (3)). 
In this case an order of magnitude change in $B$ will change $\ln V_{0}$ only by about 10\%.

According to Eq. (11), the temperature dependence of $\ln V_{0}$ may directly be estimated from 
$A(t)$ as
\begin{equation}
\ln V_0=-{{dA} \over {d\tau }}.
\end{equation}
Unfortunately, as one may see in Fig. 7, our accuracy is not sufficient to extract reliably 
the very weak temperature dependence of this parameter. At low temperatures, where the 
temperature dependence of $\ln V_{0}$ may definitely be neglected, Eq. (11) may be used to 
extract the value of $\ln V_{0}$ for the corresponding temperature range. Actually Eq. (11) 
provides two independent possibilities to evaluate $\ln V_{0}$. First, $\ln V_{0} = - dA/d\tau$ 
(Eq. (13)) and second, $\ln V_{0} = A(\tau = 0)$. Both evaluations result in the same value 
of $\ln V_{0}$, again supporting the validity of our approach. 

There is yet another way to evaluate $V_{0}$. Since at low temperatures the critical current 
of our sample is practically temperature independent (see Fig. 6), the Maley method may be 
used to establish the value of $\ln V_{0}$ in this temperature range. This has already been 
done in our previous work for the temperature range between 10 and 17 K, resulting in 
$\ln [V_{0}$ (nV)$] = 18.6$ for $H_{e} = 0$ and $\ln [V_{0}$ (nV)$] = 20.5$ for 
\begin{figure}[h]
 \begin{center}
  \epsfxsize=0.9\columnwidth \epsfbox {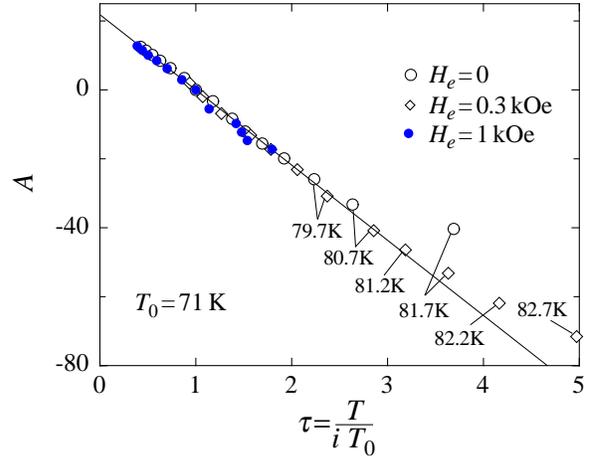}
  \caption{The parameter $A$ as a function of the normalized temperature $\tau = T/iT_{0}$ 
           with $T_{0}¥ = 71$ K. The straight line is drawn according to Eq. (11) with 
		   $\ln [V_0($nV$)]=21.8$. The temperatures related with several data points are 
		   indicated in the diagram. The data points for $H_{e}=0$ and $H_{e}=1$ kOe are 
		   shown down to $T=54$ K.}
  \end{center}
\end{figure} 
$H_{e} = 1$ kOe.$^{21}$ The straight line in Fig. 7 is drawn assuming that $\ln V_{0} = 20.5$, 
the value obtained with the Maley method for $H_{e} = 1$ kOe, while the data points shown in 
Fig. 7 were obtained by our scaling procedure, which is based on Eq. (9). As one may see in 
the inset of Fig. 7, the points for $H_{e} = 1$ kOe are very well approximated by the solid 
line up to $\tau \approx 5.5$, which corresponds to a temperature of about 50 K. Since the 
Maley method provides the value of $\ln V_{0}$ without any a priori assumptions, we consider 
this agreement as an important confirmation that Eq. (9), which is based on our two main 
assumptions, is valid. 

The deviations of the high temperature data points for $H_{e} = 1$ kOe from the straight 
line, which may be seen in Fig. 7 for $\tau \geq 6$, are most likely due to the temperature 
dependence of the attempt frequency $\nu_{0}$. In the case of $H_{e} = 1$ kOe the increase 
of $\nu_{0}$, according to Eq. (3), results in an increase of $V_{0}$ and $A(t)$ should 
deviate downward as it is indeed the case. In the case of $H_{e} = 0$ the situation is 
somewhat different. As pointed out above, in this case not only $\nu_{0}$, but also the 
magnetic induction $B$ is temperature dependent. Because $V_{0}$ is proportional to the 
product $\nu_{0}B$, a more complicated behavior of the $A(\tau)$ dependence is expected in 
this case. 

\subsection {Results for $H_{e} = 0.3$ kOe}

For $H_{e} = 0.3$ kOe the current induced by the magnetic field step was considerably smaller 
than for the other two cases, thus prohibiting reliable measurements below $T = 70$ K. For 
the scaling procedure we have chosen $T_{0}¥ = 71$ K. As a consistency check we have also 
repeated the scaling procedures for $H_{e} = 0$ and $H_{e} = 1$ kOe with this value of $T_{0}$ 
and we compare the results obtained for all three cases. 

It should be noted that in this high-temperature range the induced currents in our ring were 
rather small and the values of the magnetic induction, created by these currents at our Hall 
probe, were of the order of a few Gauss only. Such small values of the magnetic induction are 
difficult to measure accurately if the external magnetic field is high. That is why the 
measurements for $H_{e} = 0.3$ kOe could be made with higher accuracy than for $H_{e} = 1$ kOe. 
The data for $H_{e} = 0$, although accurate, are not very meaningful at high temperatures 
because of the uncertainty in the values of $B$ in the case of the remanent magnetization.

Fig. 8 shows $A(t)$ in the high temperature range. One may see that a straight line is a good 
approximation to the data up to $T \approx 81$ K. At higher temperatures, however, there are 
clear upward deviations even for the case of $H_{e} = 0.3$ kOe, which cannot be explained by 
uncertainty arguments, but rather indicate the break down of our approach. 

Fig. 8 demonstrates the good agreement between the data obtained in different fields. The 
value of $\ln [V_{0} ($nV$)] = 21.8$, estimated for this temperature range slightly exceeds 
the value of 20.5 obtained from the analysis of the low temperature data. We argue that it 
is the temperature dependence of the attempt frequency that is responsible for this 
difference. It should be pointed out, however, that for $H_{e} = 0$, the deviation of 
the points upwards starts at lower temperatures than it is the case for $H_{e} = 0.3$ kOe. 
At $H_{e} = 0$ the magnetic induction in the sample $B$ vanishes at $T = T_{c}$, which should 
result in a noticeable decrease of $\ln V_{0}$ close to $T_{c}$ (see Eq.(3)). In this case, 
according to Eq. (11), $A$ at a given value of $\tau$ should increase in agreement, 
with Fig. 8. 

The scaling procedure provides, as before, the temperature dependence of the scaling parameter 
$i(T)$, and in this case $i = I_{c}(T)/I_{c}(71K)$. For a comparison with data in Fig. 6, the 
present set $i(T)$ has to be multiplied by $I_{c}(71K)/I_{c}(13K)$. Because for $H_{e} = 0.3$ 
kOe the lowest achieved temperature was 70 K,  $I_{c}(71K)/I_{c}(13K)$ could not be evaluated 
directly. However, as one may see in Fig. 6, the difference between the $i(T)$ sets for 
$H_{e} = 0$ and $H_{e} = 1$ kOe is small. Therefore there is no risk of a significant error if 
we equate $I_{c}(71K)/I_{c}(13K)$ for $H_{e} = 0.3$ kOe with the arithmetic mean of the 
corresponding values for $H_{e} = 0$ and $H_{e} = 1$ kOe. The points calculated in this way 
are also shown in Fig. 6.

\subsection {Evaluation of the critical current and the activation energy}

As demonstrated above, the assumption that the hopping flux objects remains the same for all 
currents and temperatures together with Eq. (9) are sufficient for the scaling of the $\ln V$ 
versus $I$ curves obtained at different temperatures. In this procedure the scaling parameter 
$i$ represents the temperature dependence of the normalized critical current, but the absolute 
value of $I_{c}$ remains unknown. Below we show that the same assumptions are also sufficient 
to establish the absolute value of the critical current from the experimental data. The 
approach that we consider in this section was first used by Beasley at al..$^{28}$ At 
currents close to $I_{c}$ only a small part of the $u(x)$-function in the vicinity of 
$x = 0$ represents the essential part of the potential barrier. In this case $u(x)$ may be 
expanded in a Taylor series about the point $x = 0$. Taking into account that 
$d^{2}u/dx^{2} = 0$ at $x = 0$ and keeping only the first two non-zero terms, 
\begin{equation}
u(x)=u'_cx-bx^3.
\end{equation}
Using this analytical expression for $u(x)$, one obtains 
\begin{equation}
U(I/I_c)=\frac{4(u'_c)^{3/2}}{3\sqrt {3b}} (1-I/I_{c})^{3/2}.
\end{equation}
Hence, the current dependence of the activation energy for $(1-I/I_c)\ll 1$ should follow Eq. 
(15), independently of the particular shape of $u(x)$. In this case, one can use Eq. (15) 
together with Eq. (4) to estimate $I_{c}$, $\ln V_{0}$ and $b$ from the $V$-$I$ data. In our 
already cited previous publication the high current part of the $\ln V$ versus $I$ curve was 
fitted in this way.$^{20}$ This procedure worked reasonably well, but introducing three 
fitting parameters led to a substantial uncertainty. Now we have rather accurate estimates of 
$\ln V_{0}$, as obtained using Maley's method in Ref. 21, and therefore, we can use the same 
fitting procedure as in Ref. 20 but with only two fitting parameters. In this way we obtain 
$I_{c}(H_{e}=0) = 290$ A and $I_{c}(1$kOe$) = 301.5$ A, very similar values, as expected. 
These are the values for $T = T_{0} = 13$ K. Since $I_{c}$ is practically temperature 
independent at low temperatures, these values may safely be considered as the critical 
currents for $T = 0$. 

\begin{figure}[h]
 \begin{center}
  \epsfxsize=0.92\columnwidth \epsfbox {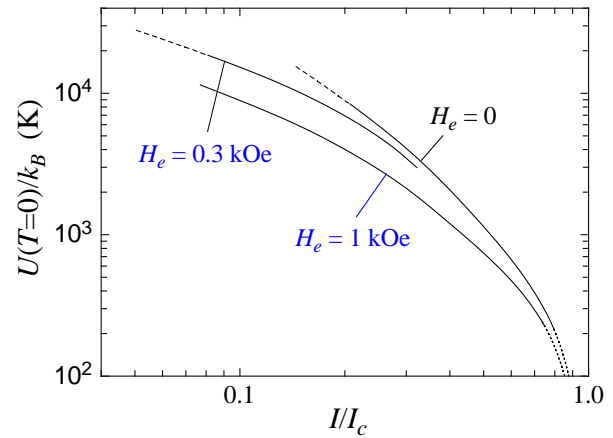}
  \caption{$U(I/I_{c}¥)$ calculated for $T = 0$. The dashed lines are calculated using data 
           from the measurements at high temperatures, where the applicability of the scaling 
		   procedure is uncertain (see text).}
 \end{center}
\end{figure}
Since $I_{c}¥(T=0)$ and $\ln V_{0}$ are now known, we may apply Eq. (4) to calculate 
$U(I/I_{c})$ from the master curves presented in Fig. 5.$^{29}$ The results are shown in 
Fig. 9 on double logarithmic scales. As mentioned above, the data at the highest temperatures 
cannot be described by our approach. Therefore the corresponding parts of the $U(I/I_{c})$ 
curves, still calculated in the same way, are indicated by the dashed lines. The different 
parts of the $U(I/I_{c})$ curves presented in Fig. 9 are calculated from the $V$-$I$ 
characteristics measured at different temperatures. As we saw, the parameter $\ln V_{0}$ is 
slightly temperature dependent, however, the exact value of $\ln V_{0}$ is only important for 
currents close to $I_{c}$ and hence low activation energies, i.e., for the analysis of 
measurements made at low temperatures. This justifies the use of the low temperature value 
of $\ln V_{0}$ for the whole temperature range. 

\subsection {Reconstruction of the shape of the potential barriers}

In our approximation, there is a direct connection between the profile of the potential 
barriers $u(x)$ in real space and $U(I/I_{c}¥)$. The function $u(x)$ may be derived from the 
$U(I/I_{c}¥)$ data as they follow from experiment. However, $u(x)$ can be found unambiguously 
only if some additional assumptions about its features are made. Here, as well as in our 
previous work,$^{20}$ we assume that the shape of the $u(x)$-function is as illustrated in 
Fig. 10(a), i.e., the point where $du/dx$ has its maximum corresponds to the bottom of the 
potential well. The somewhat more realistic potential shown in Fig. 10(b) does not alter the 
result of the calculation procedure. 

\begin{figure}[t]
 \begin{center}
  \epsfxsize=0.75\columnwidth \epsfbox {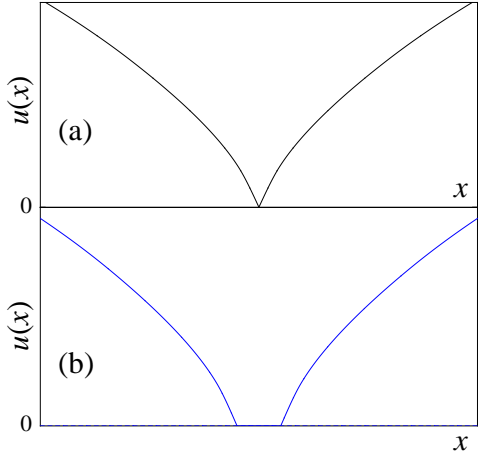}
  \caption{Schematic plots of $u(x)$ near the bottom of the potential well which have been 
           used in our calculations (see text).}
 \end{center}
\end{figure}
\begin{figure}[h]
 \begin{center}
  \epsfxsize=0.93\columnwidth \epsfbox {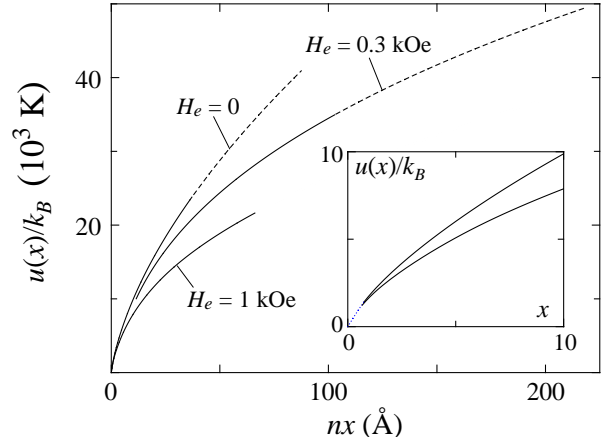}
  \caption{The potential profiles for zero current, calculated from $U(I/I_{c}¥)$. The dashed 
           lines correspond to the dashed lines in Fig. 9. The inset emphasizes the behavior 
		   of $u(x)$ for small $x$. The solid lines are calculated from the experimental data. 
		   The dotted line is an extrapolation of $u(x)$ using Eq. (14).}
 \end{center}
\end{figure}
The calculation procedure is described in detail in Ref. 20. The value of 
$u'_{c}/n \approx 2000$ K/{\AA} can be estimated from the critical current density using Eq. 
(7). The results of the calculations are presented in Fig. 11 as a function of the product 
$nx$, with $n$ being the number of the vortex lines in the hopping vortex bundle. We have 
postulated that $n$ does not depend on current and temperature and the experimental results, 
presented in this work, strongly indicate that this is indeed the case. There is no way to 
deduce $n$ directly from the experimental data. Since, however, our analysis, based on the 
fact that the hopping flux object remains the same, is rather successful for this wide range 
of temperatures, it seems most likely that we are dealing with a hopping of single vortices, 
i.e., $n = 1$. The $u(x)$-functions shown in Fig. 11 represent pinning potentials for three 
different values of the applied magnetic field. These pinning potentials include not only the 
interaction of the vortex line with one particular pinning center, but also with other 
vortices. Note that only the solid lines in Fig. 11 represent reliable results. The dashed 
lines are obtained by formally using our approach in the temperature range where its 
application is not really valid. 

\begin{figure}[h]
 \begin{center}
  \epsfxsize=0.93\columnwidth \epsfbox {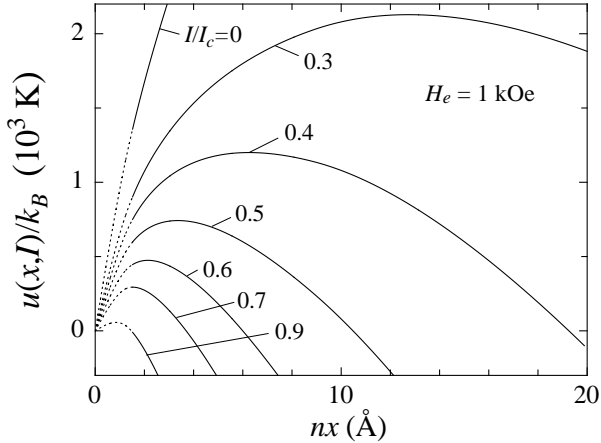}
  \caption{Calculated variations of the potential profile, shown in Fig. 11 for 
           $H_{e}¥ = 1$ kOe, with increasing current. The corresponding values of $I/I_{c}$ 
		   are indicated near the curves. The dotted lines indicate extrapolations using 
		   Eq. (14).}
 \end{center}
\end{figure} 
The electric current does not change the vortex interaction with the pinning centers or 
other vortices, but it causes a Lorentz force to act on the vortices. This force tilts the 
potential profile as is illustrated in Fig. 12. This figure clearly demonstrates that the 
position of the maximum of the potential barrier moves closer to the bottom of the potential 
well with increasing $I/I_{c}$. Fig. 13 shows the $x$ position of the maxima of $u(x,I)$ as 
a function of $I/I_{c}$. It may be seen that for most of the investigated current range, the 
extension of the potential barriers is limited to small values of $x$. 
\begin{figure}[h]
 \begin{center}
  \epsfxsize=0.9\columnwidth \epsfbox {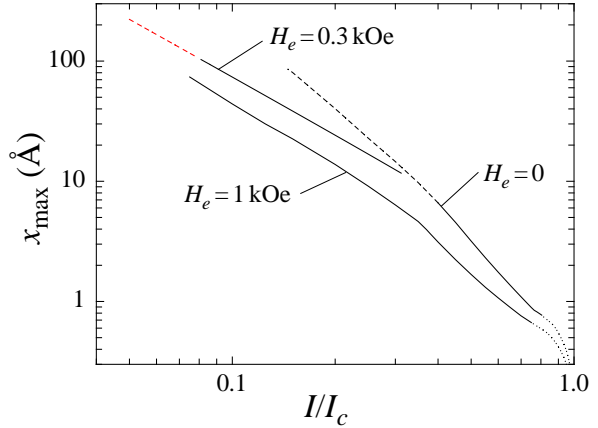}
  \caption{The position of the potential barrier maxima as a function of $I/I_{c}$. The solid, 
           dashed, and dotted lines are calculated from the corresponding curves shown in 
		   Fig. 11. Here we assume that the number $n$ of vortices in the moving flux bundle 
		   is 1 (see text).}
 \end{center}
\end{figure} 
\section {Discussion}

In this work we have applied a new scaling procedure, first described in Ref. 20, to analyze 
the experimental flux-creep data obtained for a superconducting YBCO film in a wide range of 
temperatures. The two basic assumptions on which the scaling procedure relies are (a), the 
hopping flux object remains the same for all currents and temperatures and (b), the 
temperature dependence of the flux-creep activation energy can be described by Eq. (9), 
implying that not the shape, but only the amplitude of potential barriers is temperature 
dependent. 

With these assumptions, our approach does not allow any freedom in the treatment of the 
experimental data. The two scaling parameters $i$ and $A$ of Eq. (10) and their variations 
with temperature are unambiguously determined by the shape of the experimental $V$-$I$ curves. 
It is to be noted that Eq. (11) provides the possibility to verify the consistency of the 
approach. Although the parameters $i$ and $A$ are solely evaluated by using Eq. (10), they 
should also obey Eq. (11), if our approach makes sense. As one may see in Figs. 7 and 8, the 
relation between $i$ and $A$ indeed follows Eq. (11) from the lowest investigated temperature 
of 10 K up to $T\approx 81$ K. Taking into account that in this temperature range 
$\tau = T/iT_{0}$ changes by almost a factor of 60, we consider the validity of Eq. (11) in 
this wide range of $\tau$ as unequivocal evidence that the chosen approach is meaningful and 
that Eq. (9) is indeed valid in the corresponding range of currents. A similar approach has 
successfully been applied in the analysis of the same type of $V$-$I$ data at low temperatures, 
where quantum tunneling of vortices is predominant.$^{21}$

Small deviations of the experimental points from the straight line given by Eq. (11), which 
may be seen in Fig. 8 for the highest investigated temperatures and $H_{e} = 0.3$ kOe, 
indicate that our approach is not adequate for describing the flux-creep process close to 
$T_{c}$. Taking into account the simplicity of the assumptions that have been made, this 
failure is not surprising at all. We believe that the most likely reason for these deviations 
is that Eq. (9) does not correctly describe the temperature dependence of the activation 
energy for temperatures in the vicinity of $T_{c}$. As has already been discussed, the 
experimentally available range of the $I/I_{c}$ values decreases with increasing temperature. 
It means that at high temperatures we get the low current part of $U(I/I_{c})$, which is 
mainly determined by the behavior of the $u(x)$ function far away from the bottom of the 
potential well. In other words, at high temperatures the flux-creep activation energy is 
determined by $u(x)$ at large $x$, while at low temperatures $u(x)$ at small values of $x$ 
is essential. It is not obvious that our assumption expressed in Eq. (8) is valid for large 
$x$. As we have seen, our description of the flux-creep process breaks down for $T > 81$ K 
(Fig. 8). At $T = 81$ K, the top of the potential barrier is located at a distance $x_{max}$ 
approximately 100 {\AA} away from the bottom of the potential well (Fig. 13). Comparing this 
distance with the coherence length $\xi (T=81K) \approx 50$ {\AA},$^{30}$ we may therefore 
conclude that Eq. (8) provides an adequate description of the $u(x)$-function up to 
$x \approx 2\xi (T)$. 

As has been shown in the inset of Fig. 6, $I_{c}¥(T)$ can very well be approximated by Eq. 
(12). This is a rather unexpected result. One may argue that close to $T_{c}$, where the 
Ginzburg-Landau (GL) theory is applicable, $I_{c}(T)$ should be proportional to 
$H_{cm}\xi (T) \sim (1 - T/T_{c})^{3/2}$, where $H_{cm}$ is the thermodynamic critical 
field.$^{31,32}$ This means that $I_{c}$ should vanish at $T_{c}$, which is about 2 K higher 
than $T_{dp}$. In addition the GL theory provides a different temperature dependence of 
$I_{c}$ in this regime than is dictated by Eq. (12). However, this disagreement may just as 
well be fictitious because for $H_{e}¥ = 1$ kOe we have established the $I_{c}(T)$ curve up 
to $T \approx 78$ K only, and we cannot exclude that there will be a change of the $T$ 
dependence of $I_{c}$ at higher temperatures. Although we do not have any experimental 
indication for such a change, it is important to state that our results 
do not exclude this possibility.

We now return to the temperature dependence of the normalized relaxation rate $Q$, which is 
shown in Fig. 3. In our approach all the features of the flux-creep process follow from the 
profiles of the potential barriers, which are shown in Fig. 11. Using these profiles, one may 
also calculate $Q(T)$. In an exact calculation, the temperature dependence of the critical 
current should be taken into account. But even our simplified calculation, neglecting the 
temperature dependence of $I_{c}$, gives a fairly good account of $Q(T)$, as may be seen from 
the solid line shown in Fig. 3. It thus turns out that the appearance of a plateau in 
$Q(T)$ may be traced back to a very simple shape of the potential barrier and no additional 
assumptions are needed to explain this, at first glance, very astonishing $Q(T)$ curve. This 
kind of $Q(T)$ curves is a common feature of different YBCO material, including not only 
films, but also flux-grown and melt-processed crystals.$^{12,23,24}$ The close similarity of 
the $Q(T)$ curves for all these materials leads to the natural conclusion that the plateau 
in the $Q(T)$ curves must have a common origin, implying that the profiles of the potential 
barriers in different YBCO materials are similar. There are also sufficient physical grounds 
for such a conclusion. The potential profile for a chosen pinning center is determined by the 
structure of the vortex line. The distribution of the order parameter near the vortex core 
and the distribution of the magnetic field around the vortex line are the most important 
ingredients. Because the coherence length $\xi$ and the magnetic-field penetration depth 
$\lambda$ are the relevant material parameters, it seems quite likely that the profiles of 
the potential barriers are similar in different samples of the same compound. We conclude 
that the particular combination of $\xi$ and $\lambda$ in YBCO compounds is the reason for 
the formation of a plateau in $Q(T)$. 

In this paper we have used $U(I/I_{c}¥)$ to calculate the profile of potential barriers as 
illustrated in Figs. 11 and 12. On the other hand, it is well known that HTSCÕs samples are 
not uniform and one should expect that different barriers have different shapes. In this 
situation, the physical relevance of the potential profiles calculated in the way outlined 
above, is not obvious. In order to clarify the situation, we consider the flux-creep process 
in more details. There are very many different trajectories by which the vortices are allowed 
to cross the ring sample. It is obvious, however, that only those trajectories containing the 
lowest potential barriers will actually be used. There are also many different potential 
barriers along each trajectory, but the very few with the largest amplitudes are essential in 
limiting the vortex motion. In our experiments an average value of $U/k_{B}¥T$ is 25. This 
ratio, according to Eq. (2), is related with the probability of the thermally activated 
hopping. For such large values of $U/k_{B}¥T$, even very small variations of the amplitude 
of $U$ between different barriers result in a considerable difference in the probability of 
hopping. 

In the ring geometry, the evaluation of the number $N$ of vortices which are leaving or 
entering the ring cavity per second is straightforward. Using Eq. (1) and taking into account 
that the experimentally accessible voltages range between $10^{-4}$ and 1 nV, we get $N$ 
between 50 and $5\cdot 10^{5}$ s$^{-1}$ for the lowest and the highest voltage, respectively. 
The value of $N$ for low voltages is only 50 vortex lines per second and it is difficult to 
imagine that many different trajectories are used in this case. Most likely all these vortices 
cross the sample along the easiest way and on this trajectory, only the barrier with the 
largest amplitude determines the actual flux creep rate. In our approach we assume that the 
$u(x)$-function which describes the potential profile remains the same, independently of the 
vortex transfer rate. This is why it is important to verify whether one single trajectory is 
also sufficient for transferring a much bigger number of vortices, corresponding to 
$V = 1$ nV. For $B = 1$ kG the distance between vortices is of the order of $10^{-5}$ cm. This 
implies an average vortex velocity $w \sim 5$ cm/s, if we force all $5\cdot 10^{5}$ vortices 
to follow the same trajectory across the sample in one second. This value of $w$ is rather low 
and there is no reason to expect that a single trajectory would no longer suffice for the 
transfer of vortices with increasing V in this voltage range. 

An important consequence of this line of thoughts is that the analysis of flux-creep rates 
provides information only about one particular pinning center, which represents the highest 
potential barrier for the vortex motion on the energetically most favorable trajectory across 
the sample. This is true not only for our experiments, but for all measurements of 
magnetization relaxation. It should be noted that at voltages a few orders of magnitude 
higher but still corresponding to the flux creep regime, nonlinear effects connected with the 
vortex motion may already be important. In this case one trajectory will not be sufficient 
for transferring all the vortices across the sample and, still in the flux-creep regime, a 
crossover from one to several trajectories with increasing voltage is expected. Such a 
transition is expected to be indicated by a corresponding alteration of the shape of the 
$V$-$I$ curves.

\section {Summary and conclusions}

We present a detailed experimental study of flux-creep rates in a ring-shaped superconducting 
YBCO film. A very wide range of temperatures between 10 and 83 K has been investigated. It is 
shown that all the details of the flux-creep process can be traced back to simple thermally 
activated hopping of vortices or vortex bundles over potential barriers with the hopping flux 
object remaining the same for all currents and temperatures. This is, in fact, the simplest 
possible approach for describing the flux-creep phenomenon. Using a recently developed scaling 
procedure,$^{20}$ we have succeeded in extracting the current dependence of the flux-creep 
activation energy (Fig. 9) and the temperature dependence of the critical current (Fig. 6) 
from the primary $V$-$I$ data. In the whole covered temperature range, the temperature 
dependence of the critical current $I_{c}¥(T)$ can very well be approximated by a simple 
power law (Eq. (11)). The current dependence of the activation energy $U(I/I_{c}¥)$ is then 
used to reconstruct the profiles of the potential barriers in real space (Figs. 11 and 12). 
It is important to emphasize that the outlined scaling procedure passes internal consistency 
checks and it appears that the proposed approach adequately describes the real flux-creep 
process.

In practically all previous reports where scaling procedures have been used to extract $U(I)$ 
from flux-creep data, the condition imposed by Eq. (9) has been adopted. Therefore, the main 
difference between our approach and other models is that, instead of complicated assumptions, 
we consider the simplest possible case of the vortex hopping. It is also important that we 
have chosen the shape of the experimental $\ln V$ versus $I$ curves as a criterion for 
deriving the scaling parameters. This renders our approach free from any additional 
assumptions.


\begin{references}
	
\bibitem{1} M. P. Maley, J. O. Willis, H. Lessure and M. E. McHenry, Phys. Rev. B 
{\bf 42}, 2639 (1990).

\bibitem{2} B. M. Lairson, J. Z. Sun, T. H. Geballe, M. R. Beasley, and J. C. Baravman,
Phys. Rev. B {\bf 43}, 10405 (1991).

\bibitem{3} M. E. McHenry, S. Simizu, H. Lessure, M. P. Maley, J. Y. Coulter, I. Tanaka, 
and H. Kojima, Phys. Rev. B {\bf 44}, 7614 (1991).

\bibitem{4} D. Shi and S. Salem-Sugui, Jr., Phys. Rev. B {\bf 44}, 7647 (1991).

\bibitem{5} P. J. Kung, M. P. Maley, M. E. McHenry, J. O. Willis, J. Y. Coulter, Phys. 
Rev. B {\bf 46}, 6427 (1992).

\bibitem{6} S. Sengupta, D. Shi, Z. Wang, M. E. Smith, and P. J. McGinn, Phys. Rev. B 
{\bf 47}, 5165 (1993).

\bibitem{7} S. Sengupta, D. Shi, Z. Wang, M. E. Smith, S. Salem-Sugui, Jr., and 
P. J. McGinn, Phys. Rev. B {\bf 47}, 5414 (1993).

\bibitem{8} J. R. Thompson, Y. R. Sun, L. Civale, A. P. Malozemoff, M. W. McElfresh, 
A. D. Marwick and F. Holtzberg, Phys. Rev. B {\bf 47}, 14440 (1993).

\bibitem{9} H. G. Schnack, R. Griessen, J. G. Lensink and H. H. Wen, Phys. 
Rev. B {\bf 48}, 13178 (1993).

\bibitem{10} H. Theuss and H. KronmŸller, Physica C {\bf 229},17 (1994).

\bibitem{11} S. H. Chun, S. H. Moon, Y. Chong and Z. G. Khim. Physica C {\bf 235-240}, 
2919 (1994).

\bibitem{12} H. H. Wen, Z. X. Zhao, R. J. Wijngaarden, J. Rector, B. Dam, and R. Griessen, 
Phys. Rev. B {\bf 52}, 4583 (1995).

\bibitem{13} H. H. Wen, H. G. Schnack, R. Griessen, B. Dam and J. Rector, Physica C 
{\bf 241}, 353 (1995).

\bibitem{14} J. J. Sun, B. R. Zhao, L. Li, B. Xu, J. W. Li, S. Q. Guo and B. Yin, Physica C 
{\bf 291}, 257 (1997).

\bibitem{15} Y. Yu, X. N. Xu, Z. Y. Zheng, X. Jin, and X. X. Yao, Supercond. Sci. Technol., 
{\bf 10}, 568 (1997).

\bibitem{16} H. H. Wen, P. Ziemann, H. A. Radovan, T. Herzog, Physica C {\bf 305}, 185 (1998).

\bibitem{17} J. Jung, H. Darhmaoui and H. Yan, Supercond. Sci. Technol., {\bf 11}, 
973 (1998).

\bibitem{18} E. Moratakis, M. Pissas, G. Kallias and D. Niarchos, Supercond. Sci. Thechnol. 
{\bf 12}, 682 (1999).

\bibitem{19} H. H. Wen, Z. X. Zhao, S. L. Yan, L. Fang and M. He, Physica C {\bf 312}, 
274 (1999).

\bibitem{20} I. L. Landau and H R. Ott, Physica C {\bf 331}, 1 (2000).

\bibitem{21} I. L. Landau and H. R. Ott, Physica C, in print.

\bibitem{22} I. L. Landau and H. R. Ott, Phys. Rev. B {\bf 61}, 727 (2000).

\bibitem{23} Y. Yeshurun, A. P. Malozemoff, and A. Shaulov, Rev. Mod. Phys. 
{\bf 68}, 911 (1994).

\bibitem{24} A. P. Malozemoff and M. P. A. Fisher, Phys. Rev. B {\bf 42}, 6784 (1990).

\bibitem{25} In the flux creep regime the resistance of the sample depends exponentially on 
$j/j_{c}$ and even small fluctuations of the sample cross-section or the critical current 
density reduce $L_{creep}$ significantly to below the total length of the sample.

\bibitem{26} The activation energy is a linear function of current only for triangular or 
trapezoidal potential barriers.

\bibitem{27} These error margins are errors of approximation only and do not include 
experimental errors and additional errors introduced by the scaling procedure.

\bibitem{28} M. R. Beasley, R. Labusch, W. W. Webb, Phys. Rev. {\bf 181}, 682 (1969).

\bibitem{29} For the case of $H_{e}¥ = 0.3$ kOe, $I_{c}¥(T=0)$ was estimated as an arithmetic 
mean of the corresponding values for $H_{e}¥ = 0$ and $H_{e}¥ = 1$ kOe.

\bibitem{30} G. Blatter, M. V. FeigelÕman, V. B. Geshkenbein, A. I. Larkin and V. M. Vinokur, 
Rev. Mod. Phys. {\bf 86}, 1125 (1994).

\bibitem{31} Y. Yeshurun and A. P. Malozemoff, Phys. Rev. Lett. 60, 2202(1988).

\bibitem{32} M. Tinkham, Phys. Rev. Lett. {\bf 61}, 1658 (1988).

\end{references}
\end{document}